\documentclass[manuscript=letter, utf8]{beilstein}

\usepackage{xspace}
\usepackage{textgreek}
\usepackage{amsmath}
\usepackage{amssymb}
\usepackage[sort&compress,numbers]{natbib}
\usepackage{xcolor}
\usepackage{ae,aecompl}
\usepackage{graphicx}
\usepackage{palatino}
\usepackage[T1]{fontenc}
\usepackage{setspace}
\usepackage{siunitx}
\usepackage{enumitem}

\begin{document}

\title{Direct electron beam writing of silver using a \textbeta-diketonate precursor: first insights}

\author*{Katja Höflich}{katja.hoeflich@fbh-berlin.de}
\affiliation{Ferdinand-Braun-Institut, Leibniz-Institut für Höchstfrequenztechnik (FBH) Gustav-Kirchhoff-Str. 4, 12489 Berlin, Germany}
\affiliation{EMPA Swiss Federal Laboratories for Material Science and Technology, Feuerwerkerstrasse 39, CH 3602 Thun, Switzerland}
\author[2]{Krzysztof Mackosz}
\author[2]{Chinmai S Jureddy}
\author[1]{Aleksei Tsarapkin}
\author[2]{Ivo Utke}
\maketitle

\begin{abstract}
Direct electron beam writing is a powerful tool for fabricating complex nanostructures in a single step.
The electron beam locally cleaves the molecules of an adsorbed gaseous precursor to form a deposit, similar to 3D printing but without the need for a resist or development step.
Here, we employ for the first time a silver \textbeta-diketonate precursor for focused electron beam induced deposition (FEBID). 
The used compound (hfac)AgPMe$_3$ operates at an evaporation temperature of 70 - 80°C and is compatible with commercially available gas injection systems used in any standard scanning electron microscope.
Growth of smooth 3D geometries could be demonstrated for tightly focused electron beams, albeit with low silver contents in the deposit volume.
The electron beam induced deposition proved sensitive to the irradiation conditions leading to varying compositions of the deposit and internal inhomogeneities such as the formation of a layered structure consisting of a pure silver layer at the interface to the substrate covered by a deposit layer with low silver content.
Imaging after the deposition process revealed morphological changes such as the growth of silver particles on the surface.
While these effects complicate the application for 3D printing, the unique deposit structure with a thin, compact silver film beneath the deposit body is interesting from a fundamental point of view and may offer additional opportunities for applications.
\end{abstract}


\section{Introduction}
Direct writing with an electron beam allows for single-step and maskfree 3D printing of sophisticated nanostructures at the nanoscale~\cite{Fowlkes2016, Keller2018, Winkler2019, Skoric2020}. 
The process relies on the electron beam induced fragmentation of adsorbed precursor molecules on a substrate~\cite{Wnuk2011, Huth2018, Barth2020, Yu2021, Utke2022}. 
The precursor is typically supplied in gaseous phase. 
With the different complex pathways involved in electron-induced chemistry and with different process variables involved, the composition and microstructure of the deposits can be tuned for obtaining desired nanocrystalline material~\cite{Bernau2010, Winkler2014a, Pablo-Navarro2017}. 
Such a direct writing technique is especially interesting for the fabrication of nano-optical components where the actual geometry in combination with the material composition governs the optical response of the device. However, typically a dominant carbon portion is present in the deposit due to the use of organometallic precursors~\cite{Utke2008, Utke2022} what poses practical challenges in device design fabrication. 
Accordingly, various methods to improve purity during~\cite{Langford2004, Villamor2015, Shawrav2016}  or after deposition~\cite{Botman2009, Hoeflich2011, Plank2014, Winkler2017, Reisecker2024} were developed. 
For plasmonic applications, a  metallic surface layer with a thickness exceeding the skin depth  is sufficient to obtain the desired functionality~\cite{Haverkamp2017, Hoeflich2019}. 
While pure metal deposition by direct electron beam writing was demonstrated for gold precursors with inorganic ligands~\cite{Utke2000, Mulders2012}, often high-purities come at the expense of a reduced shape fidelity~\cite{Utke2022}. 
This is also true for the recently established direct electron beam writing of silver, which demonstrated high purities of up to 74 at.$\%$~\cite{Jurczyk2023} but with large surface roughness and small vertical growth rates~\cite{Hoeflich2017, Berger2018, Hoeflich2018}. 

For silver only few solid metalorganic compounds exist that feature a sufficient vapor pressure and stability to be delivered and used into a vacuum chamber. 
Only the class  of carboxylates led to successful implementation until now, including both fluorinated and non-flourinated ligands~\cite{Jurczyk2023}. 
The surprisingly high content elemental silver that was found in the deposit despite the large number of carbon atoms in the ligands, was attributed to the thermodynamically favorable release of CO$_2$ upon ligand cleavage~\cite{Martinovic2022}. 

All successfully tested silver carboxylates exhibit a generally high reactivity and sensitivity upon electron beam impact, which shows up as pronounced deposition of halos. 
In addition, all require relatively high substrate temperatures (well above 100~\textdegree C) in order to avoid condensation.
Hence, thermal effects are expected to play an  important role in deposit shape evolution with the enhanced desorption rates contributing twofold: i) improving deposit purity due to fast desorption of cleaved ligands but also ii) decreasing the volume growth rate due to short precursor residence times. 
In addition, surface effects like enhanced dissociation due to removal of ligands by chemisorption but also an increased mobility of the metallic clusters leading to Ostwald ripening were assumed to play an important role~\cite{Utke2022}.
In order to achieve practicable vertical growth rates, new precursors are being searched for that allow lower process temperatures.

Here, we employ the compound (hfac)AgPMe$_3$ (CAS 148630-66-4, stoichiometry Ag:P:F:O:C = 1:1:6:2:8, trimethyl\-phosphine\-(hexafluoro\-acetyl\-acetonato)\-silver(I)) for focused electron beam induced deposition (FEBID).
(hfac)AgPMe$_3$ is a white to light yellow solid which was used before for chemical vapor deposition (CVD)~\cite{Dryden1993} 
and for growing silver nanoparticles by atomic layer deposition~\cite{Duan2017}. 
Like for other silver precursors a pronounced halo and a very high sensitivity with respect to electron beam impact are observed for dissociation with the weakly focused beam of a thermal electron emitter. 
However, for the first time sufficient vertical growth rates in combination with high fidelity were achieved for a tightly focused electron beam. 
The deposit composition and chemistry evolution are unexpected, leading to a structure surface decorated with silver nanoparticles and an interfacial layer of elemental silver at the bottom, the formation mechanism of which deserves further investigation. 

\section{Materials and Methods}

FEBID of (hfac)AgPMe$_3$ was carried out in a Hitachi S3600 tungsten filament scanning electron microscope (SEM). The precursor compound was evaporated using a fully integrated self-built gas injection system (GIS) consisting of chemically inert steel. No injection needles were used. The components of the GIS that were in direct contact with the precursors were cleaned after each use. The cleaning involved mechanical wiping using acetone-wetted tissues, followed by a sequence of sonification in acetone and ethanol, and dry-blowing with nitrogen. Depositions were carried out on silicon with native oxide. The silicon substrates were cleaned using a sequence of sonification in acetone, ethanol, and rinsed water, and dry-blowing with nitrogen. 
In our deposition experiments faint deposits were visible starting at a GIS temperature of 50°C for spots of 5 min dwell time turning into clearly visible deposits starting from about 60°C. From 80°C onwards a good deposition rate was observed. After several hours condensation became visible, that could be avoided by heating the substrate to a temperature of 60°C. 
All deposits shown in the following were obtained for a GIS temperature of 80°C and substrate temperature of 60°C.

Depositions with a single dwell spot duration of 5, 30, and 60~min were carried out using 15 kV primary electron energy and about 0.5~nA beam current. Rectangular patterns of 10x10 µm² were scanned in an inward spiraling beam path with a 3~nm point to point pitch, a dwell time of 1~µs per point and different number of passes using the Xenos Patterning software.

A typical workflow involved the deposition of an automated sequence of shapes during night carefully avoiding unintended electron beam impact while precursor molecules were present (cf. Supporting Information for more details). The high-resolution images presented in the main manuscript were taken in a field emitter Hitachi S-4800 (SEM).
The chemical composition of the deposits was determined using energy-dispersive X-ray spectroscopy (EDX) at a Hitachi S-4800 SEM equipped with an EDAX Genesis 4000 detector and Tescan Mira dual beam instrument with an EDAX EDX system 
.

To prove for compatibility of the tested precursor with field emitter microscopes and commercial gas-injection systems, further deposition experiments were performed using a dual beam instrument Zeiss Crossbeam 340 KMAT equipped with a commercial integrated GIS from Kleindiek based on the MM3A-EM micromanipulator platform. 
To increase the molecule supply rate a GIS version without valve was employed. 
The GIS temperature of 80°C could be stabilized and monitored with an external temperature control unit. 
Due to the specific design the crucible cap had a lower temperature (cf. SI for details).
This limited the overall available time for experiments to roughly 2 hours, since the evaporated precursor would start to crystallize at the crucible cap.
Heating of the sample was realized with a Kleindiek MHS (Micro Heating Stage), which allowed to keep the substrate temperature at constant 60° for the whole experiment.

The microstructure of the deposits was investigated using a ThermoFischer Themis 200 G3 aberration probe corrected Transmission Electron Microscope (TEM) operating at 200 kV 
. Cross-sectional TEM lamellas were prepared by standard sample preparation protocol using Tescan Lyra3 FIB-SEM system. The TEM overview image was stitched together from two individual images using the ImageJ stitching plugin https://github.com/fiji/Stitching.

\section{Results and Discussion}

\begin{figure}
\centering
\includegraphics[width=0.9\linewidth]{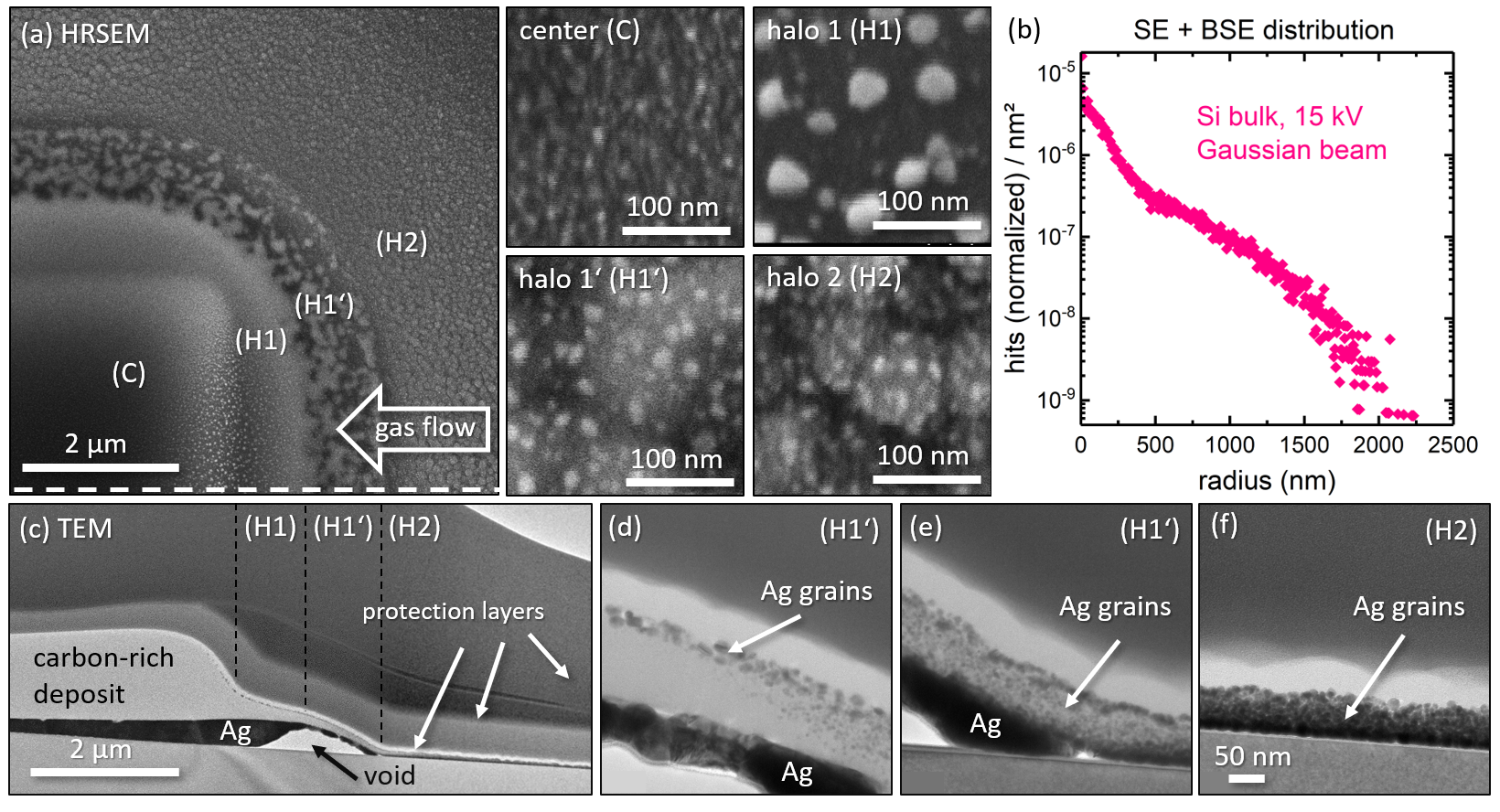}
\caption{(a) Scanning electron micrograph showing the upper right quadrant of a square deposit with the different halo regions and their close-ups and (b) the corresponding simulated range of secondary and backscattered electrons (SE $+$ BSE). The dashed line indicates the position of the cross-sectional cut for the TEM sample preparation. (c) Transmission electron micrograph of the deposit cross-section with close-ups (d) - (f). \label{fig_square}}
\end{figure}   

The scanning electron micrograph in figure \ref{fig_square} (a) shows
one quadrant of a square deposit of 10x10 \textmu m$^2$ size patterned in 
30.000 repeats. The used beam current of 0.5~nA corresponds to an electron flux of $6 x 10^{4}$ electrons per second and nm$^2$ and an overall dose of 1.7~\textmu C or $1 x 10^{17}$ electrons per nm$^2$ in the center part. 
There is a visible halo deposition region that extends several micrometers beyond the area irradiated with primary electrons and is caused by the backscattered electrons generated by the interaction with the substrate~\cite{Hoeflich2017}.
Figure~\ref{fig_square} (b) shows the corresponding Monte-Carlo simulation of the secondary and backscattered electron (SE $+$ BSE) distribution for a Gaussian beam of 250 nm FWHM impinging on a flat silicon substrate matching our deposition conditions. 
The density of SE and BSE decreases exponentially with increasing distance to the central beam and extends to roughly 2~\textmu m which corresponds to the halo region 1 (H1 and H1') observed in Fig.~\ref{fig_square} (a).

The close-up of the central region of the deposit (C) in Fig. \ref{fig_square} shows a dark appearance with uniformly sized nanoparticles of about 5-10~nm. 
When moving away from the central area, three very distinct regions for the halo morphology are found. 
The first halo region (H1) has a relatively bright appearance, with particles of about 10~-~40~nm size that are most pronounced in the direction of the molecule-delivery. 
Below, smaller particles of about 5-10~nm can be seen. 
Particles of about 5~-~20~nm sizes are observed in the second part of the first halo region (H1') where the background forms irregularly shaped dark and bright regions of several hundreds of nm in size. 
After taking the high-resolution images the imaged regions showed an increase in particle size and brightness (cf. see Supp. Info).
This points to  incomplete dissociation  in the deposition process which embeds ligand elements in the deposited material apart from the silver. 
Upon electron-irradiation such material can be further dissociated and the contained silver can form larger cluster by enhanced diffusion.  
Finally, the second halo region (H2) depicts similar particle sizes around below 20~nm sitting on top of larger bright areas of 50~–~250~nm. 
The composition of the different halo regions was studied by EDX. Due to the rich morphology observed in the deposit, no thin-film correction was applied. Instead, the EDX spectra were taken at different acceleration voltages to obtain independent data sets (see Supp. Info for an overview of all quantification results). 
The composition tendency observed was from a carbon-rich deposit in the central region to strongly increasing silver content in the halo. 
This is very similar to observations for carboxylate silver precursors~\cite{Hoeflich2017, Berger2018, Hoeflich2018, Jurczyk2023}, where the reduced electron flux in the halo region together with the substantial substrate temperature (above 100~\textdegree C) allowed for efficient desorption of the cleaved ligands instead of their further dissociation and co-deposition into the deposit. 

To further investigate the microstructure of the deposit a thin lamella along the dashed white line in Fig.~\ref{fig_square} (a) was prepared and studied by transmission electron microscopy (TEM). The TEM overview image is depicted in Fig.~\ref{fig_square} (c), its alignment and magnification was adapted to match the HRSEM image above. 
The deposit structure turned out to be extremely non-uniform with a continuous layer of elemental silver at the interface between deposit and silicon substrate. 
In addition, small elemental silver grains decorating the surface of the carbon-rich deposit were found. 
The two very different contrast regions of the first halo H1 and H1' turn out to contain a large void of about 1~\textmu m lateral extension. 
Being observable for several deposits, this void is a strong indicator of gas formation during the deposition process. 
The halo region H1 is composed of three layers: the interfacial silver layer, the carbonaceous layer and a layer of silver particles at the top.
As depicted in Fig~\ref{fig_square}(d) the upper two layers merge along H1' transitioning into the typical granular deposit structure of elemental metal particles embedded in a carbonaceous matrix sitting on top of the relatively uniform interfacial layer of elemental silver (cf. Fig~\ref{fig_square}(e)). 
The second halo region H2 depicted in Fig~\ref{fig_square}(f) becomes thinner overall and shows a high density of silver particles with a transition to continuous silver towards the bottom. 

Similar non-uniform deposit structures were observed earlier.
For pillar deposition of gold using Me$_2$Au(acac) in a water atmosphere at about 1 Pa pressure a solid metallic core surrounded by a carbon-rich shell was obtained~\cite{Molhave2003}. 
For planar deposits similar microstructures were obtained for platinum deposition using Pt($\eta^5$-CpMe)Me$_3$~\cite{Geier2014} and ruthenium deposition using (EtCp)$_2$Ru~\cite{Rohdenburg2020} both in combination with post-deposition purification employing electron beam irradiation in a water atmosphere.

The mechanism of the interfacial silver layer formation in our case may include several factors.
One factor may be the presence of water.  
In the typically used high vacuum range of$10^{-4}$ Pa in scanning electron microscopes (SEM) during deposition still a double layer of water is present on hydrophilic surfaces
This turns water into a natural co-reactant of direct electron beam writing~\cite{Barth2020} and helps in the formation of neutral volatile reaction products from the ligands, such as CO, CO$_2$,  CH$_4$ or Hhfac, removing most of the ligand elements.
A second important factor here could be the thermal energy input from the elevated stage temperature of 60°C, which increases the mobility of the formed silver atoms and clusters in the carbonaceous matrix.
Finally, collisional momentum transfer from the primary electrons  may enhance diffusion of silver and related reordering processes in the carbonaceous matrix~\cite{Utke2008}.

The clarification of the interfacial silver layer forming mechanism  deserves further in-depth studies which are beyond of the scope of this article.
This would involve surface science approaches using mass spectrometry and/or other spectroscopic techniques such as X-ray photoelectron spectroscopy, Raman or FTIR spectroscopy~\cite{Wnuk2011, Warneke2018, Yu2021} to study the nature of the desorbed and incorporated molecular fragments ideally during the irradiation process. 


Up to now, only silver pentafluoropriopionate allowed for three-dimensional growth~\cite{Hoeflich2018}. 
However, the use of this compound required relatively high stage temperatures of about 155°C (corresponding to 125°C on the sample surface) to avoid condensation. 
On the one hand side, temperature decreases the molecule residence times exponentially according to the Arrhenius law resulting in lower vertical growth rates~\cite{Utke2008}. 
On the other hand side, a distinct thermal reaction contribution is expected to kick in~\cite{Utke2022}. 
For(hfac)AgPMe$_3$ stage temperatures between 55 and 60°C were safely suppressing condensation, which improved the adsorbate residence time and in turn the vertical growth rates.
\begin{figure}
\centering
\includegraphics[width=0.5\linewidth]{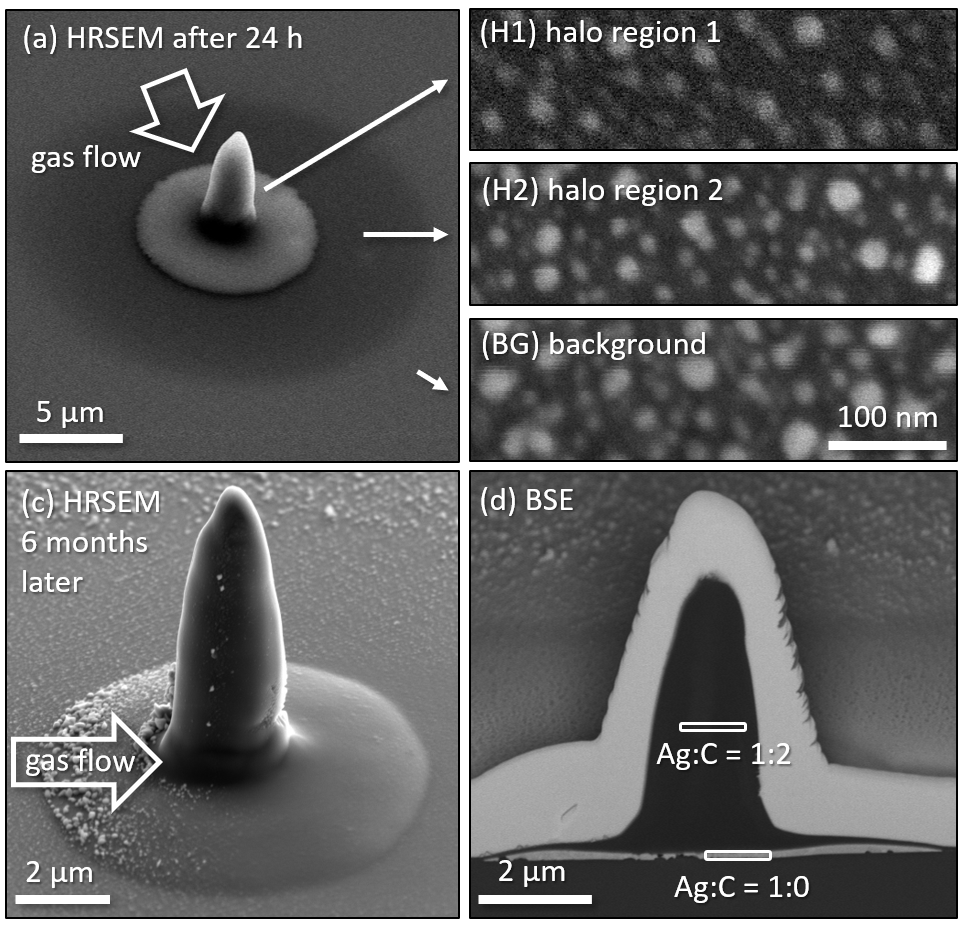}
\caption{Scanning electron micrographs of a spot deposit with 60 min continuous spot irradiation (\textbf{a}) with the corresponding close-ups of the halo regions. (c) High-resolution SEM (HRSEM) image 6 months later showing grown silver crystals in the direction of the precursor gas flow and in the further surrounding. (c) BSE image from the cross-section where heavy elements appear bright. The Ag:C ratios were obtained from EDX measurements at the cross-section in the indicated areas (CS1) and (CS2).}
\label{fig_spot}
\end{figure}   

Figure~\ref{fig_spot}(a)-(e) shows high-resolution scanning electron micrographs of a 60 min spot deposit, where the different flux regions are indicated. 
The high electron flux of about $6 x 10^{4}$ e s$^{-1}$nm$^{-2}$, leads to carbon rich 3D tip growth with a volume growth rate of 0.194\,µm$^3$/min and a vertical growth rate of about 80~nm/min. 
With continuously decreasing electron flux three different areas in the halo ca be distinguished in Fig.~\ref{fig_spot}(a). 
Here, the halo region 1 extends significantly beyond the simulated range of SE and BSE what can be attributed to forward scattering in the apex of the deposit~\cite{Bret2006}. 

EDX measurements in top view again proved an efficient release of phosphorus and fluorine in the halo region, but showed a significant amount of both in the carbon-rich deposit (cf. Supp. Information for more details).
Of note is that the surrounding of the deposit as well as the surface topography changed after deposition.
Tiny particles of few nm in appeared significantly larger after 6 months, when the cross-section was prepared. 
This again hints toward incompletely dissociated precursor that may have been further dissociated after the actual deposition process (cf. SI for more details).
The cross-sectional cuts unveils a similar microstructure as observed for the planar case. 
A carbon-rich deposit sits onto an elemental interfacial silver layer. 
Again, a tendency to form voids beneath the silver is visible.
As in the planar deposition case the silver thickness increases from the central towards the halo region.
However, for the spot exposure the pure silver layer extends significantly beyond the range of secondary and backscattered electrons. 
Hence, the silver atoms and clusters must have received sufficient kinetic energy for migration. 
The difference between planar and spot deposit is the thickness of the deposit itself.
While the silicon substrate suppresses beam-induced heating due to is good thermal conductivity, the deposit itself is most probably a bad heat conductor~\cite{Fowlkes2023}. 
Consequently, a temperature gradient with the highest temperature at the deposit surface is likely to occur~\cite{Tsarapkin2024}.
This locally increased temperature mobilizes the different species and may trigger additional thermally-driven chemical processes. 
As a result, the mobilized silver atoms/clusters would come at rest in the lowest local temperature regions at the sample surface. 


\begin{figure}
\includegraphics[width=\linewidth]{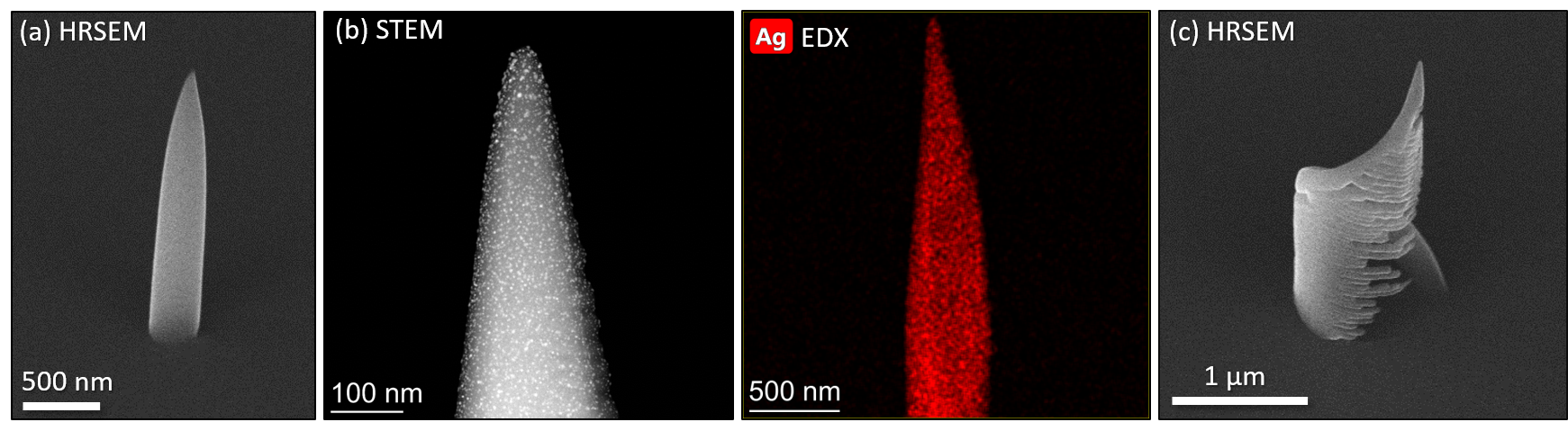}
\caption{Electron micrographs of deposition results using (hfac)AgPMe$_3$ in a commercial gas-injection system and a field emitter microscope.  (a) Pillar with smooth side walls and without noticeable halo, (b) dark-field STEM image of pillar with silver nanocrystals appearing as bright spots with corresponding EDX map, (c) helix with a radius of 500 nm and strong parasitic co-deposition below the helix arm.}
\label{fig_3D}
\end{figure}   

In a next step, the usability of (hfac)AgPMe$_3$ with a commercial gas-injection system and a field emission gun was assessed.  
In contrast to the used thermal emitter, this allows for electron beam spot sizes in the single digit nm range and an extremely increased local electron flux.
Given sufficient local precursor flux, this should lead to fast vertical growth that minimizes both, parasitic deposition in the halo region and the unwanted contribution of thermal reaction channels.


Spot deposits performed at 15~kV with 1~µs dwell time  and beam current of 246~pA.
With 8 x 10$^7$ e$^{-1}$ nm$^{-1}$ (assuming 5 nm FWHM of the beam) the local electron flux in this case was about three orders of magnitude larger compared to the case of the thermal emitter. 
The resulting pillar height was more than 2~µm.
Given the deposition time of 620~s (dose 1.5E-7 C) and the pillar width of 320 nm this corresponds to a volume growth rate of approximately 0.012 µm$^3$ per minute and a vertical growth rate of about 200 nm per minute. 

The obtained vertical growth rate was sufficient to avoid any signs of halo evolution around the structure (Fig. \ref{fig_3D} (a)). 
As this growth rate was only more the two times larger compared to the thermal emitter case, a precursor limited growth regime and thus enhanced co-dissociation of the ligands is expected.

For later TEM investigation the deposition was carried out directly on a TEM grid. 
The STEM image in Fig.~\ref{fig_3D} shows that a significant portion of nanoparticles sits on top of the deposit surface. This is in contrast to the typical deposition morphology with metal nanoparticles in a carbonaceous matrix from FEBID with organometallic precursor compounds.
The obtained silver content of about 0.6~at.\% is very low.
Apart from the large carbon portion of about 70 at.\% remarkably high contents of oxygen (22 at.\%) and phosphorous (7 at.\%) are found.
This is in-line with the favored co-deposition of ligand elements at high local electron flux.
Furthermore, compared to the initial precursor stoichiometry of Ag:P=1:1 this implies again a strong mobility of the silver which migrates from the pillar volume and enriches as an interfacial layer at the bottom.

Finally, a helix with a radius of 500~nm and one turn was deposited by scanning in a circular path with a pitch of $10^{-4}$~nm and a dwell time of 1~µs. 
These resulted in an overall deposition time of 21~min 17~s and in a beam velocity of approximately 0.15 µm per minute to match the observed vertical growth rate. 
The typical result is shown in Fig. \ref{fig_3D} (c). 
As a consequence of the large electron sensitivity, a significant parasitic co-deposition occurred below the actual helix wires caused by the residual primary electrons that penetrate the helix arms~\cite{Fowlkes2023}.
This can potentially be reduced by lowering the primary beam energy and correspondingly the interaction volume, at the same time achieving a more circular cross-section of the wires.

\section{Conclusions}
In conclusion, we studied the compound (hfac)AgPMe$_3$ as a novel precursor for focused electron beam induced deposition. 
In comparison to the available carboxylate precursors up to now, (hfac)AgPMe$_3$ is the first silver precursor that is compatible with commercial gas-injection systems as it operates at a temperature of 80\textdegree C, comparable to the ubiquituous Pt precursor Pt($\eta^5$-CpMe)Me$_3$.
At a required stage temperature of about 60\textdegree C the residence time of the molecules was long enough to realize sizable vertical growth rates and halo-free deposition for tightly focused electron beams.
While this is a promising step towards 3D printing, the ability of (hfac)AgPMe$_3$ to produce smooth vertical structures comes at the cost of the silver content.
In addition, (hfac)AgPMe$_3$ reacts extremely sensitively to the impact of electrons resulting in strong parasitic co-deposition below overhanging 3D wire structures.
Concerning the electron-induced chemistry, a strong dependence on the electron flux is revealed, leading to sharply distinguishable chemical regimes in the halo and in the deposition body. 
Here, a tendency towards incomplete dissociation at low electron fluxes and co-dissociation and deposition of the ligands at high fluxes was observed.
Together with a high mobility of the silver in the shape forming carbonaceous matrix and possible further reaction pathways, this leads to the formation of a pure silver layer at the interface to the substrate.
Although such a rich chemical evolution and the extreme sensitivity of the compound to electron impact is exciting from a fundamental point of view, it impedes the practical implementation for 3D FEBID.
One option may include the high electron flux deposition of the 3D scaffold using (hfac)AgPME$_3$ followed by low-flux irradiation to add a layer of silver crystals using the same precursor.
Finally, future research needs to identify other promising precursor candidates, ideally of the carboxylate type with long aliphatic side chains~\cite{Martinovic2022}, to combine both, stable chemistry leading to high silver contents and usability at low evaporation temperatures for 3D printing.


\begin{acknowledgements}
This work was funded by the German Research Association under grant agreement no. HO 5461/3-1 and by the Swiss National Fund by the COST-SNF project IZCOZ0\_205450. The authors want to furthermore acknowledge support by the EU COST action CA 19140 ‘FIT4NANO’ (www-fit4nano.eu).
The experiments using the field emitter microscope and the commercial GIS design were performed in the Corelab Correlative Microscopy and Spectroscopy at Helmholtz-Zentrum Berlin.
\end{acknowledgements}

\bibliography{references}
\vspace{3cm}

\newpage
\begin{suppinfo}


\subsection{Detailed EDX results}
\begin{figure}
\includegraphics[width=\linewidth]{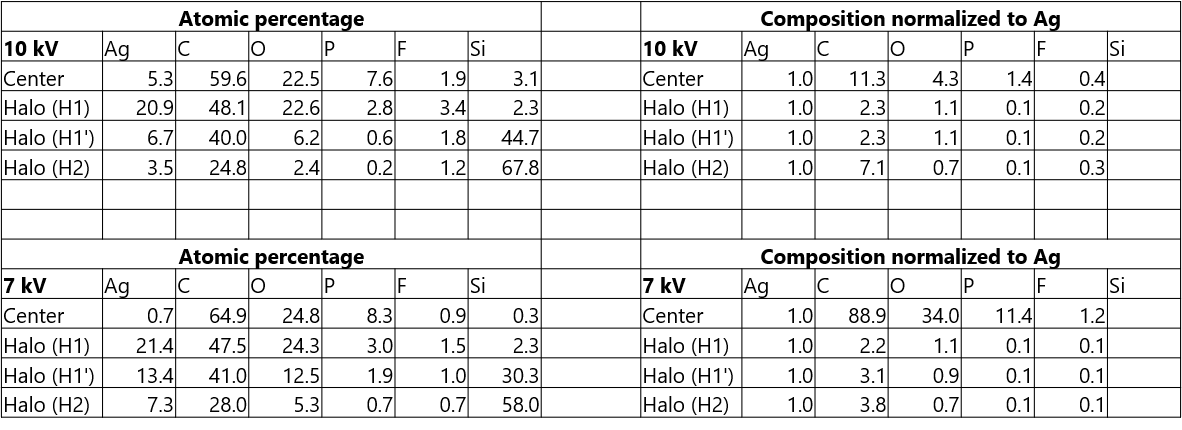}
\caption{Overview of EDX quantification results obtained for the square deposit from Figure~\ref{fig_square} in the main manuscript. In the center of the deposit the silver content is largely underestimated. We thus refer to the STEM results and the compact interfacial silver layer there. Still, the data gives a tendency to compositional changes of the carbonaceous matrix depending on the electron flux.}
\label{fig:T1}
\end{figure}  
For both types of deposits all indicated regions (central part and halo regions) were characterized by EDX. 
All of these EDX measurements were conducted in top view in the Hitachi S-4800 electron microscope, that was used for high-resolution imaging.
Acquisition of the spectra was carried out at different acceleration voltages to obtain independent data sets. 
The acquisition time of all spectra was 120 s with background spectra taken at a large field of view of about 100 µm while the actual deposit and halo spectra had a 250 nm field of view (max. magnification of 500k).
Since no thin-film correction could be applied and since the relative error of very small element quantities becomes very large, these results have to be interpreted with great caution. 
Table~\ref{fig:T1} shows the quantification results for the case of the square deposit obtained for two different acceleration voltages used for EDX where the obtained variations become immediately apparent. 
Still, we decided to add some data and discussion here, as the varying element ratios with electron dose may give useful hints towards further experiments to elucidate the actual reaction channels. 

For the rectangular deposit depicted in Figure~\ref{fig_square} the central deposit region (C) turned out to be carbon rich with with a large amount of oxygen, stable phosphorus content but strongly reduced fluorine compared to the initial precursor stoichiometry (Ag:P:F:O:C = 1:1.4:0.4:4.3:11.3 without background/layer correction). This corresponds to the layered structure with the elemental silver at the bottom and the carbonaceous deposit on top.
In halo region H1 the silver content is strongly increased in combination with an efficient removal of phosphorus, oxygen and carbon (Ag:P:F:O:C = 1:0.14:0.17:1.1:2.36 without background/layer correction). In halo region H1' apart from a substantial silver amount again more carbon is found (Ag:P:F:O:C = 1:0.08:0.26:0.9:5.9 without background/layer correction). 
In halo region H2 the oxygen content is further reduced while the carbon content is increased (Ag:P:F:O:C = 1:0.07:0.34:0.7:7.1 without background/layer correction).

\begin{figure}
\includegraphics[width=\linewidth]{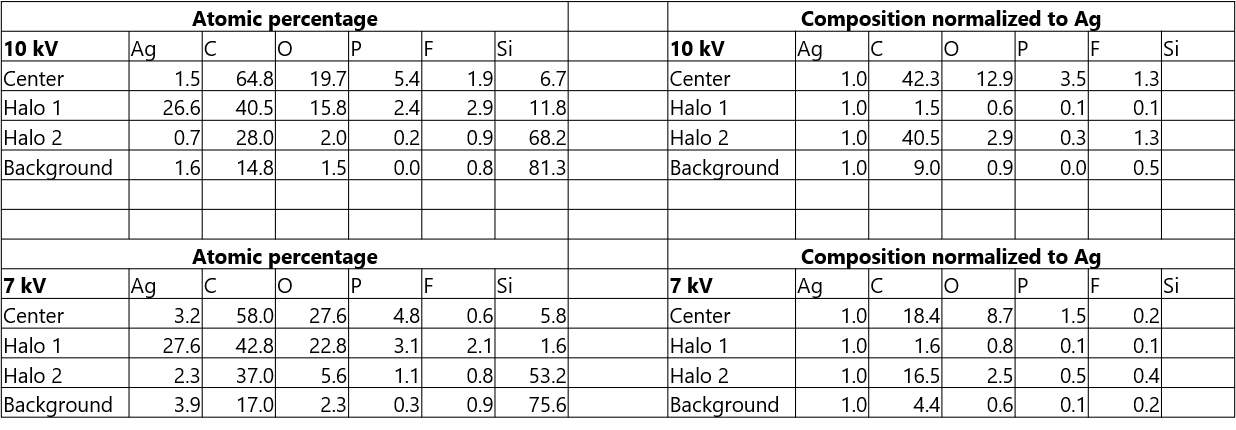}
\caption{Overview of EDX quantification results obtained for the 60 min spot deposit from Figure~\ref{fig_spot} in the main manuscript. In the center of the deposit the silver content is largely underestimated. We thus refer to the HRSEM images and EDX quantification of the cross-section and the compact interfacial silver layer there. Still, the data gives a tendency to compositional changes of the carbonaceous matrix depending on the electron flux.}
\label{fig:T2}
\end{figure}  

For the spot deposit in the central part even less silver is found. The carbon-rich deposit again released most of the fluorine (Ag:P:F:O:C = 1:4:1:9-13:34-43 without background correction). The fact that per one  atom of phosphorus only one fluorine atom but 2 oxygen and 8-10 carbon atoms are found  (stoichiometry 1:6:2:8 in the molecule), i.e. all oxygen and carbon of the ligands but mostly no fluorine, hints towards specific reaction channels during the co-deposition of the hfac ligand that deserve further investigation. When moving to a lower electron flux regime (halo region 1), the phosphorus and carbon were efficiently released during deposition as well, resulting in high silver contents of about 30 at.\% (Ag:P:F:O:C = 1:0.1:0.1:0.5:1.5 without background correction). For further decreasing electron flux (halo region 2) the carbon and phosphorus and fluorine content increased again leading to an approximate composition of Ag:P:F:O:C = 1:0.3:1.3:3:40 without background correction (Attention: large error here due to small absolute values!) The third region is close to the background with no detectable traces of phosphorus and a composition of (Ag:P:F:O:C = 1:0:0.47:1:9 without background correction).

The common trend here is a certain low-electron flux regime that allows for efficient removal of most of the ligands. 

\subsection{Deposit evolution}

The following sequence of images is intended to show the high sensitivity of the precursor to electron beam impact and the rich morphology evolution obtained after growth.

All low-resolution images were taken in-situ in the deposition SEM with a tungsten cathode. The high-resolution images were taken in a field emitter SEM.
As the precursor is extremely sensitive to electron beam irradiation any unintended irradiation was avoided.
This was realized by automated patterning of a sequence of deposits which were only imaged hours after the GIS and sample had been cooled down and the base pressure was recovered.
Figures~\ref{fig:S1} (a) and (b) depict such a deposition sequence of 4 squares with increasing number of repeats (100, 200, 1000, 30000) followed by a series of 5~min dots to assess a potential drop of the precursor flow over the several hours of deposition time. The feature on the very left side was deposited for focusing.
\newpage%

\begin{figure}
\centering
\includegraphics[width=0.89\linewidth]{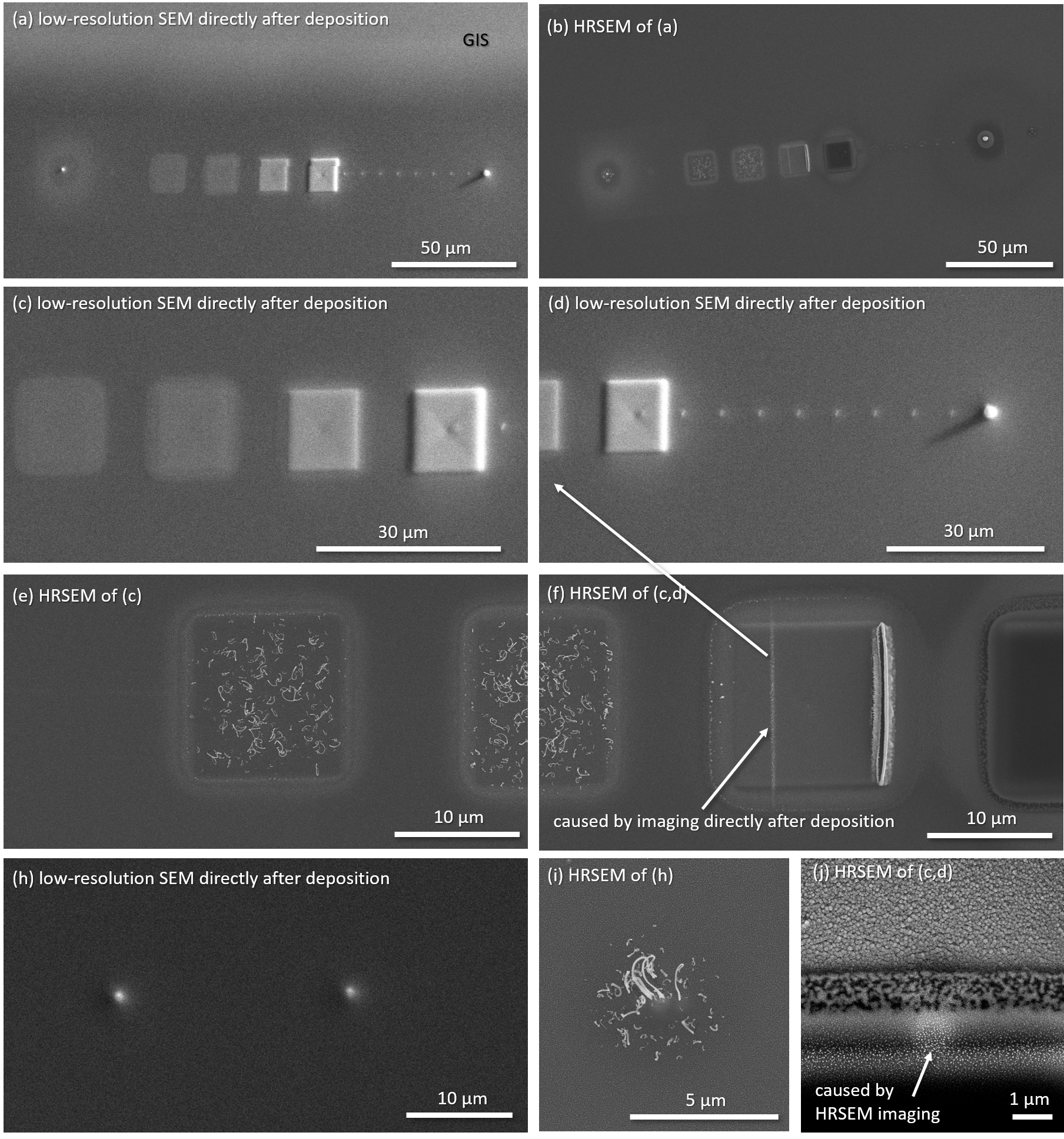}
\caption{Scanning electron micrographs of deposits directly taken after deposition in the tungsten emitter low-resolution system compared images taken with high resolution in a field emitter microscope about 12-24 hours later.}
\label{fig:S1}
\end{figure}
One immediate and recurring observation was a very smooth deposit appearance directly after deposition but additional features later being observable in high-resolution imaging.%

Figures~\ref{fig:S1} (c) and (d) show close-ups taken in the deposition SEM.
Even though the resolution is limited here, our assumption is that the filament like structures visible in Figures~\ref{fig:S1} (e) and (f) evolved only later from uncovered silver crystals at the deposit surface. 
This assumption is supported by the fact, that such features were not obtained for very long deposition times where silver migration to the interface was dominant.
In case of remaining incompletely dissociated precursor also the in-situ imaging itself could have an effect as this is equivalent to (very short) low-electron flux deposition.
To rule out this possibility as much as possible, the deposition experiments were run during night. 
The GIS heating was switched off in the early morning, the stage heating one hour later, and the in-situ images were then only taken once the base pressure was recovered.

Still, Figure~\ref{fig:S1} (f) shows a sign of this imaging by the vertical line of silver crystals that corresponds to the image frame of Figure~\ref{fig:S1} (d).
The reason here could be remaining precursor molecules adsorbed (chemisorbed) at the surface, as stated in the main manuscript, that are incompletely dissociated.

The evolution of filaments was also present for the case of the 5~min spot deposits. 
Such spot deposits are our paradigm test and focusing feature and depicted in an in-situ close-up in Figure~\ref{fig:S1} (h).
Figure~\ref{fig:S1} (i) in high-resolution imaging shows the complete reorganization of the deposit by forming filament-like structures with the longest filaments arising from the deposit center (supposedly the largest silver reservoir) but also short filaments arising from the halo region.
We cannot speculate on the specific reaction pathways nor even the actual point in time where these reactions took place.
We can only state that must have happened somewhen after the actual deposition process and before the high-resolution imaging.
Hence, we leave this observation to further investigations.
Finally, Figure~\ref{fig:S1} (j) shows the halo region of the studied square deposit with an enhanced contrast. 
Here, it becomes apparent that also later electron-impact induces slight modification which provides another hint towards incomplete dissociation.

\begin{figure}
\centering
\includegraphics[width=0.8\linewidth]{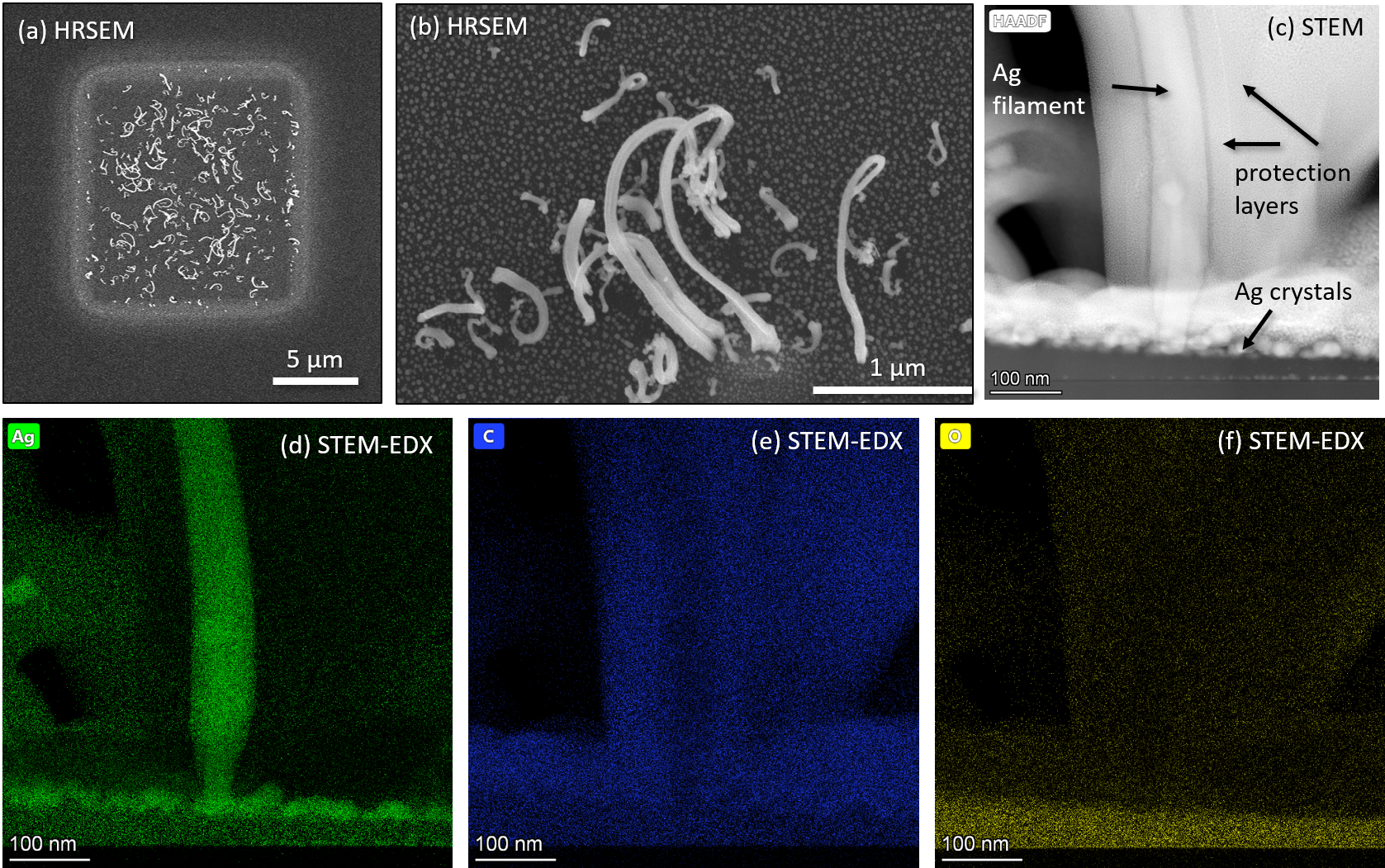}
\caption{Morphology evolution. (a) and (b) HRSEM images of filaments that become visible at later inspection but were not present directly after deposition. (c) STEM HHADF image of a lamella prepared through one filament along with the corresponding EDX maps (d)-(f) which prove that the filament is silver}
\label{fig:S2}
\end{figure}   

The obtained filaments were further studied in STEM and EDX.
Here, the preparation of electron-transparent lamellas was extremely challenging.
Still, one filament could be isolated and prepared and STEM-EDX could prove for the presence of silver.
Interestingly, the bottom crystal from which the filament grew shows a smaller diameter than the filament itself.
The results support the assumption that the filaments are from silver grown from particles present at the deposit surface but after the actual deposition process.
This could rely on reorganization of silver clusters due to their high mobility but also on thermally driven reactions of the incompletely dissociated precursor.

\begin{figure}
\centering
\includegraphics[width=0.9\linewidth]{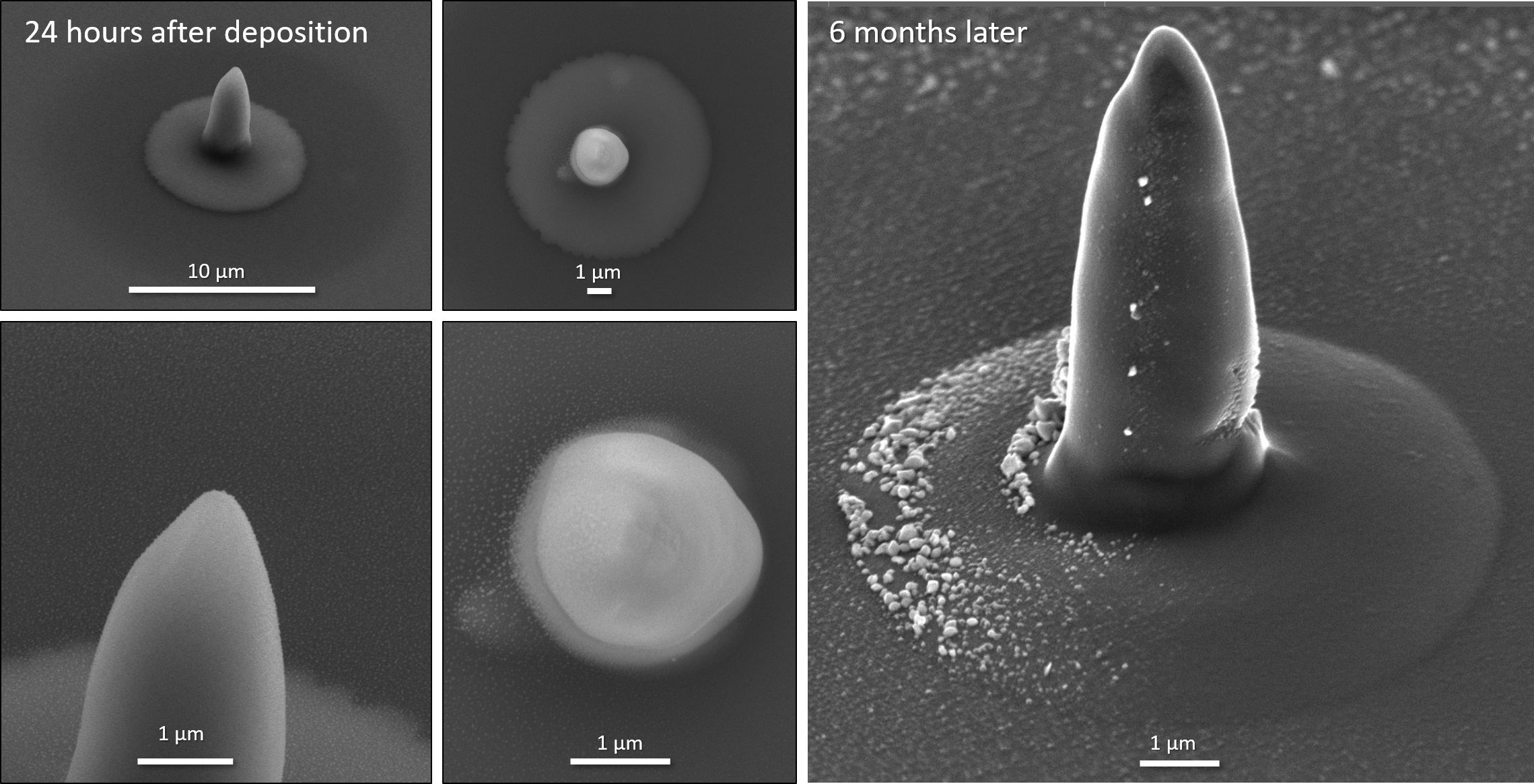}
\caption{High resolution scanning electron micrographs of the 60 min spot deposit taken 24 hours after deposition. The sample was taken out of the deposition SEM and imaged in the field emitter SEM. On the right side this is compared to a HRSEM image taken directly before the preparation of the cross-section 6 months later.}
\label{fig:S3}
\end{figure}

The assumption that the mobility of silver and its reorganization plays an important role can be further supported by the time evolution of the 60~min spot deposit. 
Here, the first high-resolution SEM images were taken about 24 hours after the actual deposition process. 
Figure~\ref{fig:S3} depicts the spot deposit with close-up in top and tilted view. 
In the direction where the GIS was located, tiny particles are visible in halo region 1. 
On the top surface of the actual deposit even smaller particles are visible.
After the surprising TEM results of the square deposits, a cross-section was prepared for this sample as well, 6 months after the actual deposition process and the imaging.
The HRSEM image taken directly before the preparation is shown on the right hand side of Fig.~\ref{fig:S3} and depicts a significant growth of the particles.
This growth was not an immediate result imaging, which existed (cf. Fig.~\ref{fig:S1} (j)) but were minor compared to the morphological changes obtained here.
Again, we cannot clarify based on the available data the exact reaction mechanisms here.

\end{suppinfo}


\end{document}